\documentclass[journal]{IEEEtran}
\usepackage{graphicx}
\usepackage{caption}
\usepackage{subcaption}
\usepackage{float}

\newcommand{\fig}[1]{Fig.~\ref{#1}}

\begin{document}
\title{Performance of the New Amplifier-Shaper-Discriminator Chip for the ATLAS MDT Chambers at the HL-LHC}


\author{\underline{H.~Kroha$^{*1}$}\thanks{$^*$Corresponding author: kroha@mpp.mpg.de}, S.~Abovyan$^1$, A.~Baschirotto$^2$, V.~Danielyan$^1$, M.~Fras$^1$, F.~M\"uller$^1$, S.~Nowak$^1$, F.~Resta$^2$, M.~De~Matteis$^2$, R.~Richter$^1$, K.~Schmidt-Sommerfeld$^1$, Y.~Zhao$^1$
  \\ \textit{$^1$Max-Planck-Institut f\"ur Physik, Munich}
  \\ \textit{$^2$Univ. of Milano-Bicocca, Physics Department, Milano, Italy}}

\maketitle
\pagestyle{empty}
\thispagestyle{empty}

\begin{abstract}
	The Phase-II Upgrade of the ATLAS Muon Detector requires new electronics for the readout of the MDT drift tubes.
	The first processing stage, the Amplifier-Shaper-Discriminator (ASD), determines the performance of the readout for crucial parameters like time resolution, gain uniformity, efficiency and noise rejection.
	An 8-channel ASD chip, using the IBM 130~nm CMOS 8RF-DM technology, has been designed, produced and tested.
	The area of the chip is 2.2~x~2.9~mm$^2$.
	We present results of detailed measurements as well as a comparision with simulation results of the chip behaviour at three different levels of detail. 
\end{abstract}

The HL-LHC at CERN will operate at peak luminosities factors of 5 – 7.5 beyond the original LHC design value of 10$^{34}$~cm$^{-2}$s$^{-1}$.
The high luminosity is a challenge for the readout system of the Monitored Drift Tube chambers (MDT) in the ATLAS Muon Spectrometer~\cite{ATLAS} in two respects. Higher hit rates, mainly due to increased cavern background, drive data transmission to the rear end electronics to the limit of available bandwidth.
In addition, the new operating parameters of the L1 trigger - latency up to 60~$\mu$s and trigger rates up to 400~kHz - call for a replacement of the entire readout chain of the MDT chambers.

In this process of renewing the MDT readout, particular attention must be given to the first stage of the readout chain, the 
Amplifier with Shaping network and Discriminator (ASD)~\cite{ASD}.
This stage determines critical quantities, like signal risetime, signal-to-noise performance and threshold uniformity among the 8 channels of the chip, which are decisive for system parameters like spatial resolution of the track coordinates (represented by the drift time in the MDT tubes) and tracking 
efficiency. 

\begin{figure}[h!]
	\centering
	\includegraphics[width=0.5\textwidth]{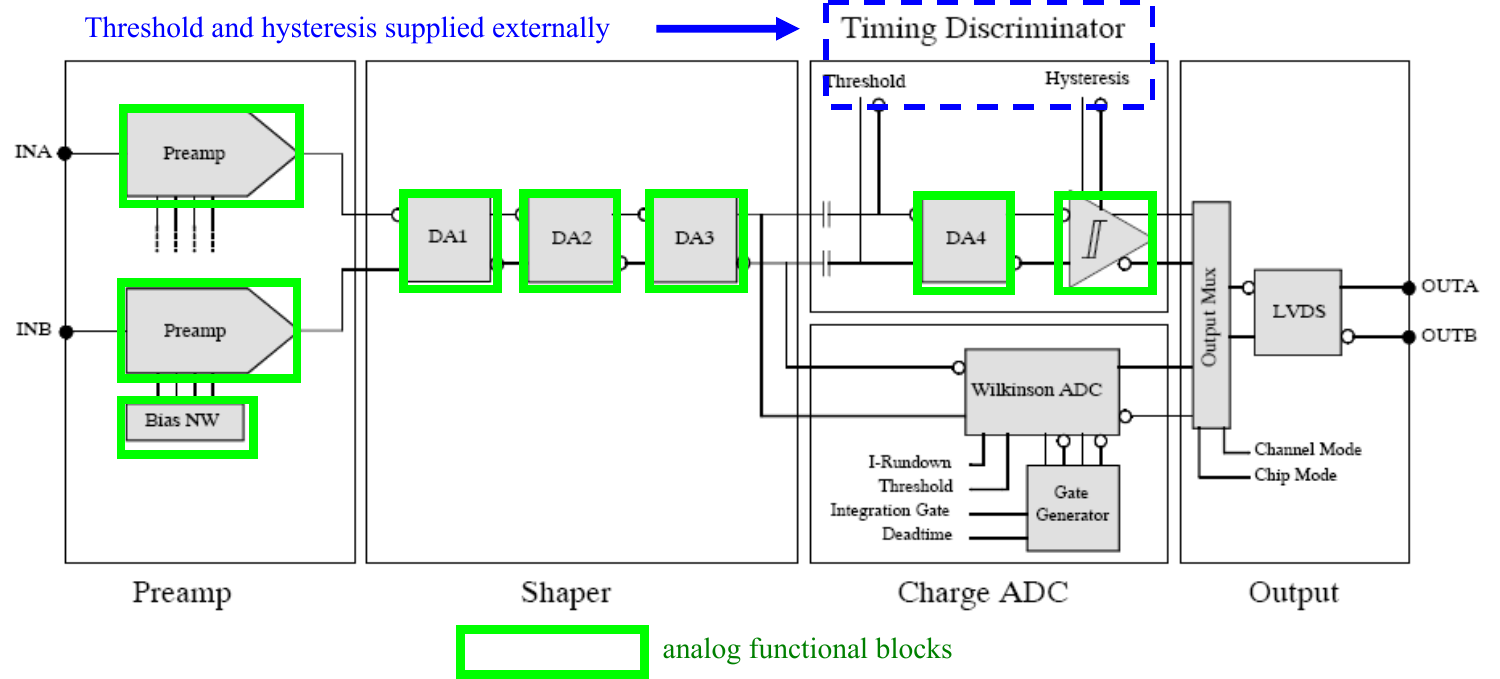}
	\caption{The functional block diagram of the ASD.}
	\label{fig::block}
\end{figure}

\begin{figure}[th]
	\centering
	\begin{subfigure}[b]{0.5\textwidth}
		\centering
		\includegraphics[width=0.9\textwidth]{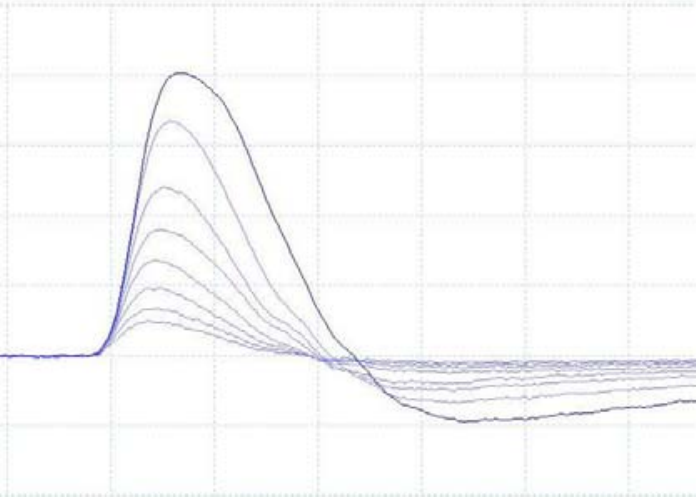}
		\caption{}
		\label{fig::delta1}
	\end{subfigure}
	\begin{subfigure}[b]{0.5\textwidth}
		\centering
		\includegraphics[width=0.9\textwidth]{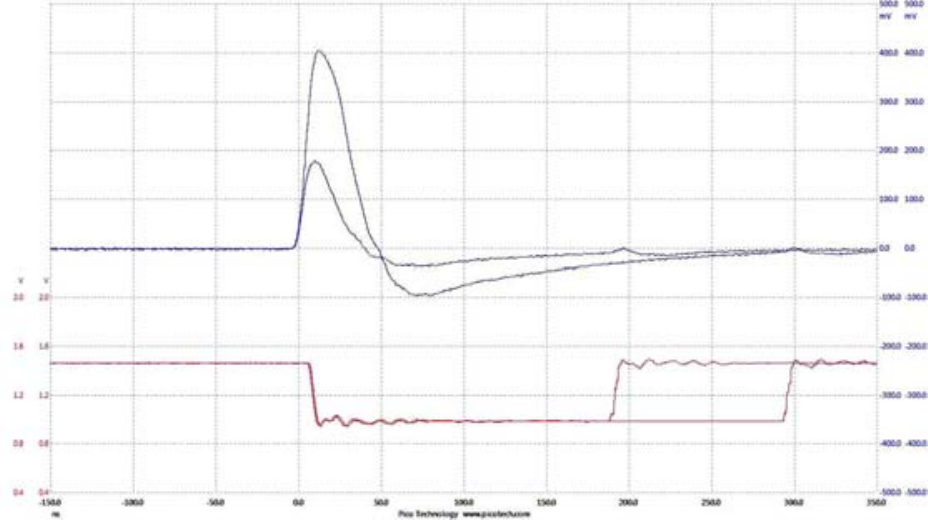}
		\caption{}
		\label{fig::delta2}
	\end{subfigure}
	\caption{(\subref{fig::delta1}) Delta pulse response of the ASD. Horiz/vert.scale: 20 ns/100 mV. (\subref{fig::delta1}) Shaped signal at two different
amplitudes and corresponding output in ADC mode. The length of the output is a measure of the charge of the pulse amplitude. Horiz./vert. scale: 50~ns/ blue: 100~mV,red: 400~mV.}
	\label{fig::deltas1}
\end{figure}

To cope with these requirements, a chip was developed in the IBM 130~nm CMOS 8RF-DM technology.
The design contains a preamplifier, three shaping stages and a discriminator (see \fig{fig::block}). 
The chip can be operated in two output modes, the time-over-threshold (ToT) and the ADC mode.
In the ADC mode, implemented as a Wilkinson ADC, the time elapsed between leading and trailing edge is proportional to the the charge inside a predefined integration window of about 15 ns, which is an approximate measure of the amplitude of the initial signal triggering the discriminator.
The ADC information allows to apply a slewing correction to the time of threshold crossing (measurement of the drift time) and is useful for monitoring the stability of the gas gain and other operational parameters over extended periods.
The size of this 8-channel chip is 2.2 x 2,9 mm$^2$.

\begin{figure}[t]
	\centering
	\includegraphics[width=0.5\textwidth]{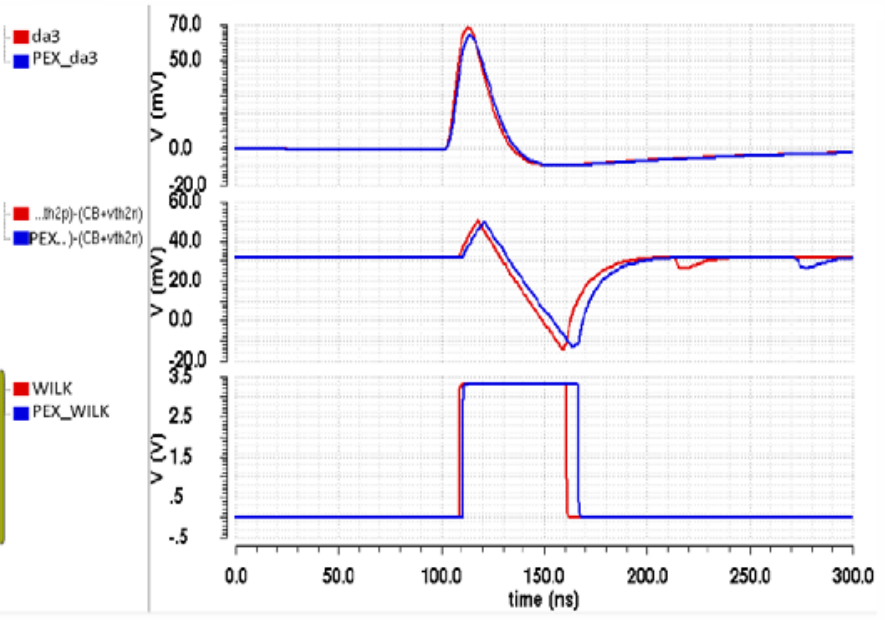}
	\caption{Simulated delta pulse response of the New ASD and ADC response.}
	\label{fig::delta3}
	
        \centering
	\includegraphics[width=0.5\textwidth]{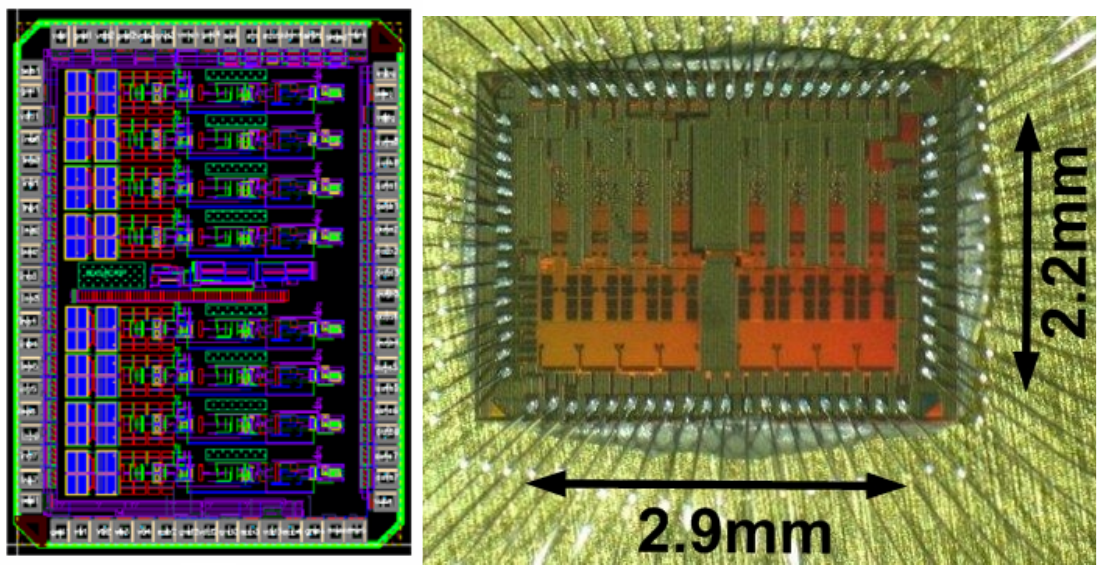}
	\caption{Simulated delta pulse response of the New ASD and ADC response.}
	\label{fig::layout}
\
\end{figure}

We present measurements of crucial performance parameters of the chip like signal rise time, signal-to-noise, uniformity of signal gain and threshold among the 8 channels.
The hysteresis of the discriminator is presented as well as the linearity of the ADC w.r.t. the incoming signal.
As an  example, the measured pulse shape behind the third shaping stage (D3 in \fig{fig::block}) is shown in \fig{fig::delta1}.
The \fig{fig::delta2} shows the same signal at two amplitudes together with the output response in the ADC mode.

The measured quantities are compared to the results of detailed simulation work, done before the submission of the chip.
\fig{fig::delta3} shows the simulated pulse shape (top) and the resulting output in ADC mode (bottom).
The middle track shows the voltage on the internal storage capacitor, where the charge in the integration window is stored at the beginning of the cycle. 
\fig{fig::layout} shows the layout of the 8-channel chip 
and the internal structure of the chip, respectively.

The critical influence of stray capacitances on the peaking time of the signal will be discussed in more detail in the presentation, demonstrating how details of the preamplifier 
layout may be decisive for the performance of the final chip.

\end{document}